\documentclass[conference]{IEEEtran}
\IEEEoverridecommandlockouts
\usepackage{cite}
\usepackage{amsmath,amssymb,amsfonts}
\usepackage{algorithmic}
\usepackage{graphicx}
\usepackage{textcomp}
\usepackage{xcolor}
\usepackage{enumitem}
\usepackage{bm} 
\usepackage{algorithmic}
\usepackage{algorithm}
\usepackage{mathrsfs}

\usepackage{makecell}  
\usepackage{amsmath}   
\usepackage{graphicx}   
\usepackage{subcaption} 

\usepackage[top=1.7cm, bottom=1.7cm, left=1.6cm, right=1.6cm]{geometry}

\usepackage{stfloats}
\newcommand{\figureref}[1]{Fig.~\ref{#1}}
\def\BibTeX{{\rm B\kern-.05em{\sc i\kern-.025em b}\kern-.08em
    T\kern-.1667em\lower.7ex\hbox{E}\kern-.125emX}}
\begin{document}

\title{Preconditioned Conjugate Gradient for MIMO-AFDM System\\
\thanks{This paper is supported in part by National Natural Science Foundation of China Program(62271316, 62101322), National Key R\&D Project of China (2019YFB1802703), Shanghai Key Laboratory of Digital Media Processing (STCSM 18DZ2270700) and the Fundamental Research Funds for the Central Universities.

The authors: Jun Zhu, Yin Xu, Dazhi He, Haoyang Li, YunFeng Guan and Wenjun Zhang are from Shanghai Jiao Tong University. Dazhi He is from Pengcheng Laboratory. The corresponding author is Dazhi He (e-mail: hedazhi@sjtu.edu.cn).
}
}

\author{
    \IEEEauthorblockN{Jun Zhu\IEEEauthorrefmark{1}, Yin Xu\IEEEauthorrefmark{1}, Dazhi He\IEEEauthorrefmark{1}\IEEEauthorrefmark{3}, Haoyang Li\IEEEauthorrefmark{1}, Yunfeng Guan\IEEEauthorrefmark{2}, Wenjun Zhang\IEEEauthorrefmark{1}, \textit{Fellow, IEEE},\\  }
    \IEEEauthorblockA{\IEEEauthorrefmark{1} Cooperative Medianet Innovation Center (CMIC), Shanghai Jiao Tong University\\\IEEEauthorrefmark{3}Pengcheng Laboratory \\Shanghai 200240, China \\ Email: \{zhujun\_22,  xuyin, hedazhi,  lihaoyang,  zhangwenjun\}@sjtu.edu.cn\\\IEEEauthorrefmark{2} Institute of Wireless Communication Technology, College of Electronic Information and Electrical Engineering \\ Email: yfguan69@sjtu.edu.cn}
}

\maketitle

\begin{abstract}Affine frequency division multiplexing (AFDM) is a promising chirp-assisted multicarrier waveform for future high mobility communications. A significant challenge in MIMO-AFDM systems is the multi-user interference (MUI), which can be effectively addressed by employing precoding techniques. However, the complexity introduced by AFDM makes the precoding process computationally expensive and challenging. To overcome this issue, We combine AFDM channel sparse property and using Preconditioned Conjugate Gradient (PCG) method to iteratively process the precoding, thereby reducing the complexity of the precoding design. Simulation results demonstrate that the proposed sparsification approach, coupled with the PCG method, achieving  quite precoding performance while significantly reducing computational complexity. This makes the application of AFDM more feasible and efficient for high-mobility communication scenarios, paving the way for its broader implementation in next-generation communication systems.

\end{abstract}

\begin{IEEEkeywords}
afﬁne frequency division multiplexing (AFDM), precoding, multiple-input multiple-output (MIMO), preconditioned conjugate gradient (PCG)
\end{IEEEkeywords}

\section{Introduction}
\IEEEPARstart{N}ext-generation wireless systems and standards, such as those beyond 5G and 6G, will provide a wide range of advanced data services. These services include ultra-reliable, high-speed, low-cost, and low-latency communications in highly dynamic environments. Examples include vehicle-to-everything, unmanned aerial vehicles (UAVs), and high-speed trains. In this context, fast time-varying channels pose significant challenges \cite{b1}, \cite{b2}). Conventional multicarrier systems mainly rely on Orthogonal Frequency Division Multiplexing (OFDM), which has the advantages of efficient hardware implementation and robustness against inter-symbol interference under linear time-varying conditions. However, in fast time-varying channels, OFDM is highly sensitive to carrier frequency offsets and Doppler frequency shifts/spreads, which often lead to severe degradation of error performance.

Recently, affine frequency division multiplexing (AFDM) has emerged as an attractive solution for efficient and reliable communication over high-mobility channels \cite{b3}. In AFDM, information symbols are multiplexed over multiple orthogonal chirped subcarriers via the discrete affine Fourier transform (DAFT) \cite{b4}. By optimizing the chirp parameters according to the channel delay-Doppler profile, AFDM achieves a separable quasi-static channel representation, enabling full diversity in doubly selective channels \cite{b3}. Notably, the DAFT used in AFDM generalizes several other transforms, including the discrete Fourier transform (DFT) and discrete Fresnel transform (DFnT), making AFDM a superset of OFDM in certain configurations. Moreover, AFDM maintains high compatibility with conventional OFDM systems, as the DAFT can be efficiently implemented using fast Fourier transform (FFT) operations with only two additional single-tap filters \cite{b5}. These advantages position AFDM as a promising multicarrier waveform candidate for next-generation wireless communication systems.

On the other hand, the convergence of MIMO and AFDM, MIMO-AFDM can be used to achieve higher spectral efficiency and lower error rate. Although channel estimation for MIMO-AFDM was studied in \cite{b6}, precoding design in MIMO-AFDM is largely an open research topic. Multi-user interference (MUI) under MIMO-AFDM systems is a problem that needs to be solved and precoding techniques can be a good solution to this problem. However, AFDM makes the complexity of precoding increase dramatically, which makes precoding techniques will no longer be advantageous. In order to solve this problem, the use of PCG technique makes it possible to solve the MUI problem under MIMO-AFDM.

In \cite{b7}, they modified this method to compute factorized sparse approximate inverses $L_M$ where the sparsity pattern of the approximate inverse is also captured automatically similar to the SPAI algorithm. In \cite{b8}, they propose a novel algorithm for the Factorized Sparse Approximate Inverse (FSAI) computation that makes use of the concept of supernode borrowed from sparse LU factorizations and direct methods. PCG of \cite{b7} and \cite{b8} solve sparse matrices, which reduces complexity compared to traditional linear precoding and non-linear coding.

 The rest of the paper is composed as follows. In Section \text{II}, the MIMO-AFDM model is introduced. In Section \text{III}, PCG is introduced and the complexity overhead of the proposed precoding is analyzed. In Section \text{IV}, simulation results show that AFDM is  robust to DFO and the proposed algorithm has lower complexity than MIMO-OFDM. In Section \text{V}, the conclusions of this paper are summarized.
 
Notation: Bold uppercase letters denote matrices and bold lowercase letters denote vectors. For a matrix $\textbf A$, $\textbf A^T$, $\textbf A^H$, $\textbf A^{-1}$, and Tr($\textbf A$) denote its transpose, conjugate transpose, inverse, and trace, respectively. blkdiag($\textbf H_1, ..., \textbf H_K$) denotes a block diagonal matrix with $\textbf H_1, ..., \textbf H_K$ being its diagonal blocks. The space of M × N complex matrices
is expressed as $\mathbb{C}^{M\times N}$.

\section{SYSTEM MODEL}

\begin{figure}[t]
\centering
\includegraphics[width=0.5\textwidth]{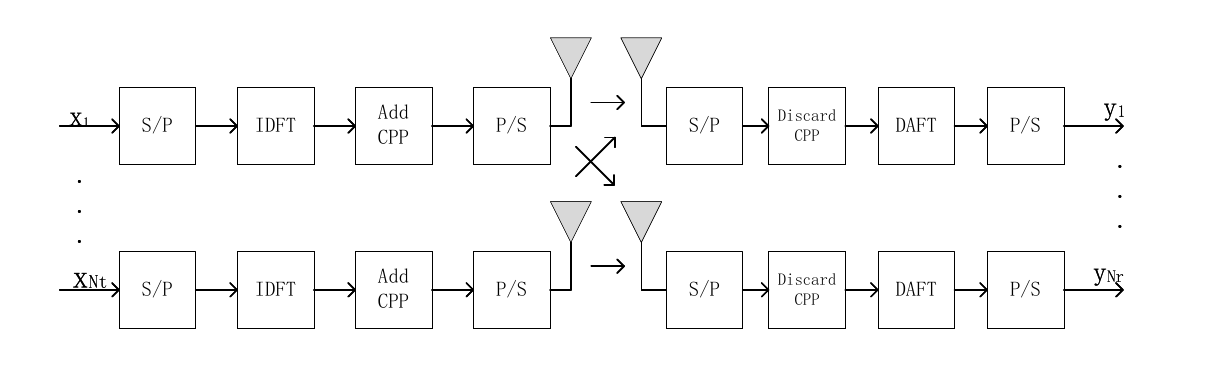}
\caption{ MIMO-AFDM of Scenario.}
\label{fig:1}
\end{figure}

In this section, we introduce the system model of MIMO-AFDM, including MIMO and AFDM. The MIMO-AFDM of scenario is shown in \figureref{fig:1}.
\subsection{AFDM Channel Model}

The AFDM modulation scheme is originally based on DAFT, such that the AFDM signal vector $\textbf s \in {\mathbb{C}^{{N}\times1}}$ ($N$ being the number of chirp carriers) can be expressed as
\begin{equation}
\begin{aligned}
\textbf s = \mathbf{\Lambda }_{c_1}^H \textbf F^H \mathbf{\Lambda }_{c_2}^H\textbf x=\textbf A\textbf x,
         \end{aligned}
\label{eq:1}
\end{equation}
\begin{figure*}[!hb]
	\hrulefill
	\vspace*{5pt}
\begin{align}
     y(m) = \frac{1}{N}\sum_{i=1}^{P}\sum_{m'=1}^{N-1} h_i e^{j\frac{2\pi}{N}(Nc_1l_i^2-m'li+c_2(m'^2-m^2))\frac{e^{-j2\pi(m+ind_i-m'+\beta_i)-1}}{e^{-j\frac{2\pi}{N}(m+ind_i-m'+\beta_i)-1}}} x(m')+N(m).  
\tag{2}  
\label{eq:2}
\end{align}
\begin{align}
   SINR_{k} = \frac{|{\textbf h_{k}}^H\textbf F_{k}|^2}{{|{\textbf h_{k}}^H\textbf F_{i}|^2 }+{\sum_{j\neq k}^{K}|{\textbf h_{j}}^H\textbf F_{j}|^2}+\sigma^2} 
\tag{21}  
\label{eq:21}.
\end{align}
\end{figure*}
where $\mathbf{\Lambda_c}\triangleq diag(e^{-j2\pi cn^2},n = 0, 1, ..., N-1)$, $c_1$ and $c_2$ are two AFDM parameters, $\textbf F$ is the DFT matrix with $e^{-j2\pi mn/N}/\sqrt{N}$, $\textbf A = \mathbf{\Lambda_{c_1}^H} \textbf F^H \mathbf{\Lambda_{c_2}^H} {\mathbb{C}^{{N}\times N}} $represents the DAFT matrix, $\textbf x \in \mathbb{A}^{N \times 1}$ denotes a vector of $N$ quadrature amplitude modulation (QAM) symbols that reside on the DAFT domain, $\mathbb{A}$ represents the modulation alphabet. We have the input-output relationship of SISO-AFDM system in the DAFT domain as shown in \eqref{eq:2} \cite{b9} at the bottom of the next page. For the convenience of illustration, we define

\begin{equation}
\begin{aligned}
 c(l_i,m,m') = e^{j\frac{2\pi}{N}(Nc_1l_i^2-m'l_i+c_2(m'^2-m^2))}
         \end{aligned}
         \tag{3}  
\label{eq:3},
\end{equation}
and the spreading factor caused by fractional Doppler as
\begin{equation}
\begin{aligned}
&F(l_i,v_i,m,m') = \frac{e^{-j2\pi(m+ind_i-m'+\beta_i)-1}}{e^{-j\frac{2\pi}{N}(m+ind_i-m'+\beta_i)-1}}\\
&ind_i = (\alpha_i+2Nc_1l_i)_N
         \end{aligned}
         \tag{4}  
\label{eq:4},
\end{equation}
where non-negative integer $l_i \in[0, l_{max}]$ is the
associated delay normalized with $T_s$ normalization, $m $ denotes DAFT domains indices, $N$ denotes the number of subcarriers (chirps), $v_i = \alpha_i + \beta_i$ represents the associated Doppler shift normalized with subcarrier spacing and has a finite support bounded by $[-v_{max}, v_{max}]$, $ind_i$ denotes the index indicator of the i-th path, $ \alpha_i\in [-\alpha_{max}, \alpha_{max}]$ and $\beta_i \in (-\frac{1}{2} , \frac{1}{2} ]$ are the integer and fractional parts of $v_i$ respectively, $v_{max}$ denotes the maximum Doppler and $\alpha_{max}$ denotes its integer component. In this paper, we assume that $l_{max}$ and $ v_{max}$ are known in advance.  

According to \eqref{eq:2}\cite{b9}, the channel model of MIMO-AFDM. Where $N_t$ and $N_r$ represent the number of transmit antennas (TA) and receive antennas (RA), respectively.
\begin{equation}
\begin{aligned}
 y_r(m) =& \sum_{t=1}^{N_t}\sum_{i=1}^{P}\sum_{m'=1}^{N-1} \frac{1}{N}h_i^{(r,t)}c(l_i,m,m') \\
         &\quad \times F(l_i,v_i,m,m')x_t(m)+W_r(m),
         \end{aligned}
         \tag{5}  
\label{eq:5}
\end{equation}
where  $P$ is the number of paths, $t\in{[0,N_t]}$ denote the index of the RA and TA respectively, $h_i^{(r,t)}$ is the channel gain of the $i-$th path between the $r-$th RA and the $t-$th TA, $W_r \in \mathcal{CN}(0, \sigma_k^2)$ represents the noise in DAFT domain at the $r-$th RA.

\eqref{eq:5} can be written as
\begin{equation}
\begin{aligned}
\textbf y_1 =  &\textbf H_{1,1}\textbf x_{1,1}+\textbf H_{1,2}\textbf x_{1,2}+\dots+\textbf H_{1,N_t}\textbf x_{1,N_t}+\textbf w_1\\
 &\vdots\\
\textbf y_{N_r} = & \textbf H_{N_r,1}\textbf x_{N_r,1}+\textbf H_{N_r,2}\textbf x_{N_r,2}+\dots+\textbf H_{N_r,N_t}\textbf x_{N_r,N_t}\\&+\textbf w_{N_r}
         \end{aligned}
         \tag{6}  
\label{eq:6},
\end{equation}

\begin{equation}
\begin{aligned}
\textbf H_{r,t} = \sum_{i=1}^Ph_i^{(r,t)}\textbf H_i
         \end{aligned}
         \tag{7}  
\label{eq:7},
\end{equation}

\begin{equation}
\begin{aligned}
\textbf H_i = \frac{1}{N}c(l_i,m,m')F(l_i,v_i,m,m')
         \end{aligned}
         \tag{8}  
\label{eq:8},
\end{equation}
where $r\in{[0,N_r]}$ and $t\in{[0,N_t]}$.

We define the effective MIMO channel matrix for the above MIMO-AFDM system as
\begin{equation}
\begin{aligned}
   \textbf H_{MIMO} = 
\begin{bmatrix}
 \textbf H_{1,1}&\dots&& \textbf H_{1,N_t}  \\
\vdots&\ddots \\
 \textbf H_{N_r,1}&\dots&& \textbf H_{Nr,N_t} \\
\end{bmatrix}.
\end{aligned}
\tag{9}  
\label{eq:9}
\end{equation}

That \eqref{eq:6} can be rewritten as
\begin{equation}
\begin{aligned}
\textbf y_{MIMO} = \textbf H_{MIMO}\textbf x_{MIMO}+\textbf w_{MIMO}
         \end{aligned},
         \tag{10}  
\label{eq:10}
\end{equation}
where $\textbf y_{MIMO}=[\textbf y_1^T, \textbf y_2^T,\dots, \textbf y_{N_r}^T]^T\in{\mathbb{C}^{NN_r\times 1}}$, $\textbf H_{MIMO}\in{\mathbb{C}^{NN_r\times NN_t}}$, $\textbf x_{MIMO}=[\textbf x_1^T, \textbf x_2^T,\dots, \textbf x_{N_t}^T]^T\in{\mathbb{C}^{NN_t\times 1}}$, $\textbf w_{MIMO}=[\textbf w_1^T, \textbf w_2^T,\dots, \textbf w_{N_r}^T]^T\in{\mathbb{C}^{NN_r\times 1}}$.

\section{Preconditioned Conjugate Gradient}
In the MIMO-AFDM system, directly precoding the channel matrix $\textbf H$ would incur prohibitively high computational overhead. To address this issue, we leverage the inherent sparsity of the AFDM channel to first sparsify $\textbf H$, followed by preconditioned conjugate gradient (PCG)-based precoding. This approach significantly reduces computational complexity while effectively mitigating multi-user interference (MUI) in MIMO-AFDM.
\begin{figure}[t]
\centering
\begin{minipage}{0.35\textwidth}
    \centering
    \includegraphics[width=\linewidth]{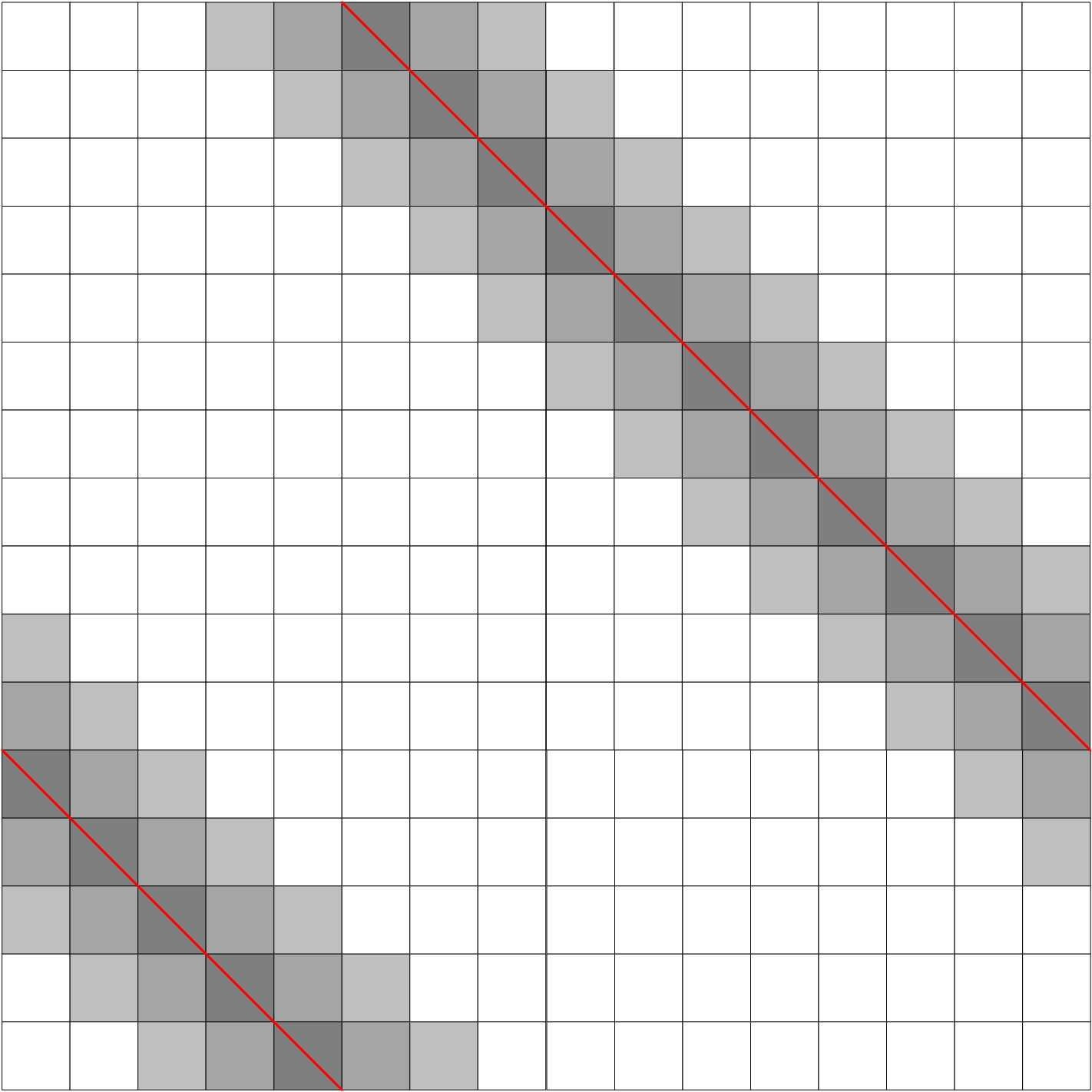}
    \subcaption{Message propagation of a one-path channel.}
    \label{fig:2a}
\end{minipage}
\hfill
\begin{minipage}{0.35\textwidth}
    \centering
    \includegraphics[width=\linewidth]{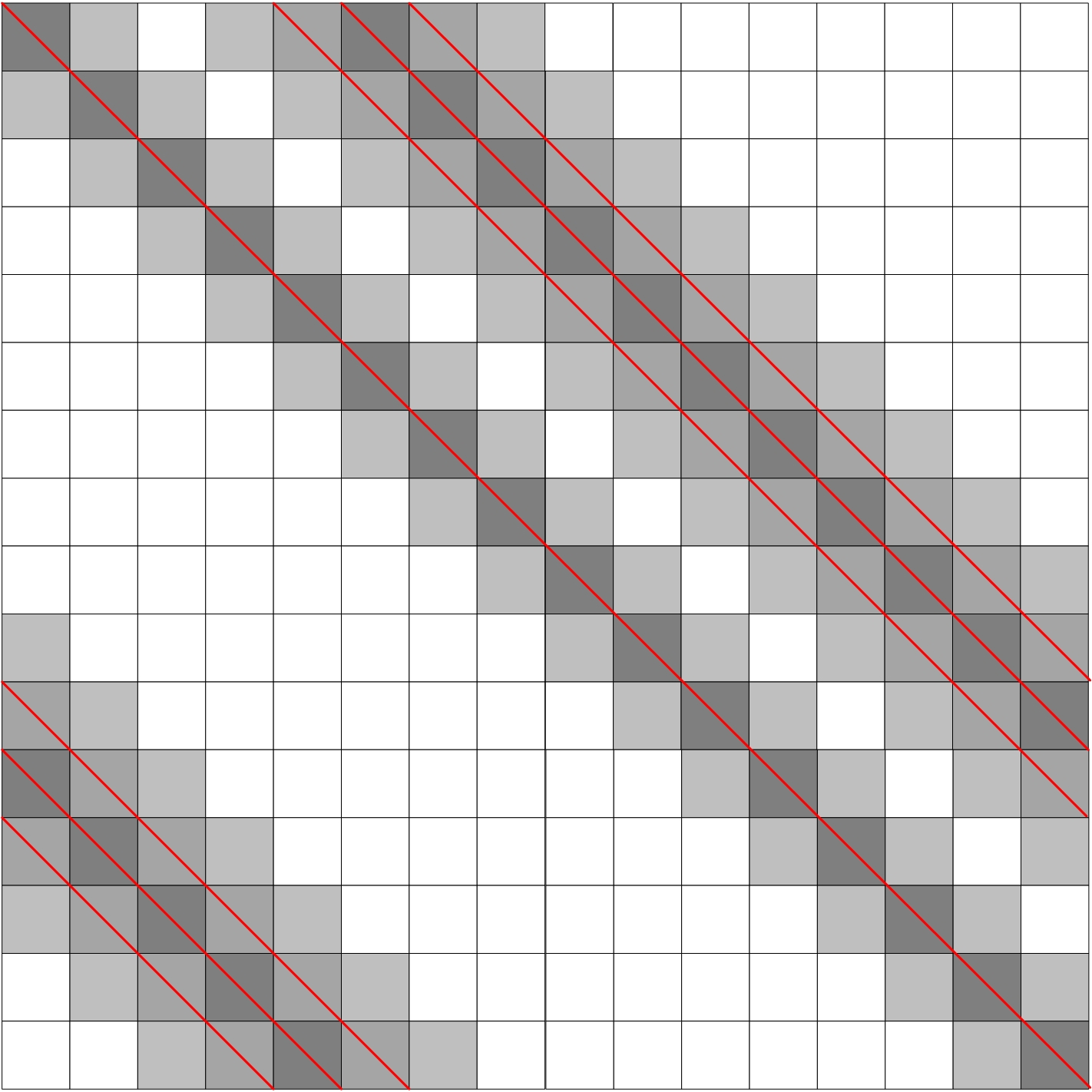}
    \subcaption{Message propagation of a two-path channel.}
    \label{fig:2b}
\end{minipage}
\caption{An example of the selected entries, highlighted by the red lines.}
\label{fig:2}
\end{figure}
\subsection{Sparse Channel Based on eSNR}
For the PCG used in this paper, the computational complexity depends on the number of elements of the channel $\textbf H$. For MIMO-AFDM systems, one possible place to reduce the computational complexity is to reduce the number of elements of $\textbf H$, thus reducing the number of unwanted message propagation. This is justified because some non-zero elements of $\textbf H$ are of very small magnitude and messages propagated on the corresponding edges may not significantly change the error rate performance. For this reason, we propose the eSNR sparse channel approach to determine the nonzero elements used for message propagation.

As shown in Fig. 2, the single and double diameter cases are represented, and N=16 is used as an example. We need to sparsify H according to Eq. 9, assuming that there are $NN_t $rows and $NN_r$ columns, and we need to judge eSNR for each element. eSNR represents the threshold value, and the element at the $s$-th cyclic shift will be added to $\textbf D_{(n, r)}$ for message propagation when and only when the eSNR of the $s$-th cyclic shift is larger than the threshold value,  

\begin{equation}
\begin{aligned}
 eSNR_s = 10log\frac{G_s}{N_0}\geq eSNR_{th},0\leq N-1,
         \end{aligned},
         \tag{11}  
\label{eq:11}
\end{equation}

where $eSNR_{th}$ denotes the threshold and eSNR denotes the eSNR at the $s-$th cyclic shift. \figureref{fig:2} illustrates an example of selected entries for message propagation for one-path and two-path channel with $N = 16$. Element at $s=5$ in \figureref{fig:2a} and elements at $s = 0, 4, 5$  and $6$ cyclic shifts in \figureref{fig:2b} are added to $D_{(n, r)}$ for message propagation.
\subsection{Preconditioned Conjugate Gradient}
The sparsification of $\textbf H$ in the previous subsection makes sense for the iterative algorithm, making the computational complexity overhead lower. We aim to solve the following problems
\begin{equation}
\begin{aligned}
 \textbf A \textbf x= \textbf b
         \end{aligned},
         \tag{12}  
\label{eq:12}
\end{equation}
where $\textbf A$ denotes the channel matrix $\textbf H$ and $\textbf b$ denotes the original data. In order to use the lower complexity iterative algorithm PCG, it is obtained that
\begin{equation}
\begin{aligned}
 \textbf f_x =\frac{1}{2}\textbf x^H\textbf A \textbf x-\textbf x^H\textbf b
         \end{aligned}.
         \tag{13}  
\label{eq:13}
\end{equation}

We can get the first order derivative and the second order derivative

\begin{equation}
\begin{aligned}
 \frac{d}{dx}\textbf f_x =\textbf A \textbf x-\textbf b
         \end{aligned}.
         \tag{14}  
\label{eq:14}
\end{equation}

\begin{equation}
\begin{aligned}
\frac{d^2}{dx^2}  \textbf f_x =\textbf A 
         \end{aligned}.
         \tag{15}  
\label{eq:15}
\end{equation}

We let $P_0$ be the negative gradient at $x_0$
\begin{equation}
\begin{aligned}
\textbf P_0 =\textbf b-\textbf A \textbf x_0
         \end{aligned}.
         \tag{16}  
\label{eq:16}
\end{equation}

Then the residual at moment $t$ is expressed as
\begin{equation}
\begin{aligned}
\textbf r_t =\textbf b-\textbf A \textbf x_t
         \end{aligned}.
         \tag{17}  
\label{eq:17}
\end{equation}

The gradient $\textbf P$ at moment $t$ can be obtained as
\begin{equation}
\begin{aligned}
\textbf P_t =\textbf r_t-\sum_{i<k}\frac{\textbf P_i^H\textbf A \textbf r_t}{\textbf P_i^H\textbf A \textbf P_i}\textbf P_i
         \end{aligned}.
         \tag{18}  
\label{eq:18}
\end{equation}

\begin{equation}
\begin{aligned}
 \alpha_t =\frac{\textbf P_t^H(\textbf b-\textbf A \textbf x_t)}{\textbf P_t^H\textbf A \textbf P_i}=\frac{\textbf P_t^H \textbf r_t}{\textbf P_t^H\textbf A \textbf P_i}
         \end{aligned}.
         \tag{19}  
\label{eq:19}
\end{equation}

Finally, we can get $x$ at the $t+1$ moment  
\begin{equation}
\begin{aligned}
 \textbf x_{t+1} =\textbf x_{t}+\alpha_t\textbf P_t
         \end{aligned}.
         \tag{20}  
\label{eq:20}
\end{equation}

According to eSNR the values below the threshold in the channel $\textbf H$ can be rounded off, which will greatly reduce the number of elements in the channel $\textbf H$, as can be seen in \figureref{fig:2}. Meanwhile, in MIMO-AFDM systems, each subcarrier of AFDM traverses all bandwidths, and the traditional precoding algorithm ZF is related to the inverse of channel H, which makes it a catastrophic problem for the precoding design, however, with sparse channel $\textbf H$, the computational complexity overhead of the PCG is only related to the number of elements and the number of iterations, so sparsification of the channel matrix H for the PCG is very much necessary. Driving a good precoding design under MIMO-AFDM.

Considering that the data symbols of each user are i.i.d. and follow a Gaussian distributed, the instantaneous uplink  $SINR_k$ of user $k$ regarding subarray $s$ can be defined as \eqref{eq:21}.
\begin{algorithm}[]
    \caption{ Efficiently Precoding in XL-MIMO-AFDM System}
	\label{algorithm1} 
	\renewcommand{\algorithmicrequire}{\textbf{Input:}}
	\renewcommand{\algorithmicensure}{\textbf{Output:}}
	\begin{algorithmic}[1]
		\REQUIRE The inverse of the SNR $\xi \geq 0$, the channel matrix of user j is $\textbf{H}_j$, and the number of algorithm iterations $T_P$;              
		\ENSURE $\textbf x_{t+1}$ from \eqref{eq:20}.
             \STATE Get $\textbf H_{r,t},$ from \eqref{eq:7}.  
     \FOR{$idex1=1$ to $N_r$}
         \FOR{$idex2 = 1$ to $N_{t}$}
       \STATE $\textbf H = \textbf H_{r,t}(:,:,idx1,idx2)$;
                 \IF{$\textbf H == 0$}
                   \STATE break;
                  \ENDIF
             \FOR{$idex3=1$ to $N$}
                \FOR{$idex4=1$ to $N$}
               \STATE $eSNR = 10log \frac{\textbf H(:,:,idx3,idx4)}{SNR}$ is form \eqref{eq:11};
                 \IF{$eSNR > -12$}
                   \STATE  $\textbf D{(idx3, idx4)}$= $\textbf H(idx3,idx4)$;
                 \ENDIF
                 
                \ENDFOR
             \ENDFOR           
               \STATE $\textbf A \textbf D = \textbf b$ is from \eqref{eq:12};
               \STATE  $r_0$ is from \eqref{eq:16};
               \STATE  $d = r_0$;
                \STATE $r_{old} = r_0^Hr_0$;
                \FOR {$idex5 = 1$ to $T_{P}$}
                   \STATE $r_{new} = r_{old} $;
                   \IF{$sqrt(r_{new})<err$}
                  \STATE $\textbf {PCG}_{out} =\textbf H\textbf x +\textbf N$;
                        \STATE break;
                        \ELSE
                        \STATE $A_p =\textbf D d$;
                         \STATE $ alpha = d'r / (d' A_p);$  
       \STATE $x = x + alpha * d$;  
        \STATE$ r = s -  H * x$;  
       \STATE $rsnew = r' * r$;  
       \STATE       $\textbf {PCG}_{out} = 0 $;
       \STATE       $d = r + (rsnew / rsold) * d$;  
       \STATE    $rsold = rsnew$;  
                    \ENDIF
                    
                \ENDFOR
               \ENDFOR

            \ENDFOR
             \RETURN  $\textbf x$
	\end{algorithmic} 
\end{algorithm}
\subsection{Complexity Analysis}
According to the proposed rKA and SwoR-rKA algorithm in \cite{b12} shows higher performance in the system and likewise faces lower complexity than ZF. 

 In this subsection, we analyze and compare the complexity of  ZF, rKA, SwoR-rKA and PCG respectively, by means of floating point operations (FLOPS) computation, \ $\textbf H\textbf H^H, \textbf H\in {\mathbb{C}^{{n}\times m}}$ requires $n^2m$ FLOPS. 
Assume that $N_t=N=n$. Combining to classic ZF needs $2n^2m$ FLOPS and rKA needs $N_{ts}T_s$ FLOPS, $T_s$ represents the number of iterations of rKA.
\begin{table}[h]
\begin{center}  
   \caption{Computational Complexity for Precoding Methods based on Complex Operations.}  
    \renewcommand{\arraystretch}{2.5}
    \begin{tabular}{c|c|c}  
        \hline
        \hline  
        \textbf{Algorithm} & \textbf{FLOPS} & \textbf{XL-MIMO-AFDM} \\   
        \hline  
        ZF & $N^22K^2N_{ts}$ & $32768N^2 $\\  
        rKA & $N_{ts}*T_s$ & $12800N^2 $ \\  
        SwoR-rKA & \makecell{$  N_{ts} \times T_s$ \\ $+ 2N_{ts} \times K$} & \makecell{$16896N^21$} \\  
        PCG & \makecell{$ nnz{\textbf H} +nnz{\textbf H}\times T_p$} & \makecell{$2816N^2$} \\ 
        \hline  
    \end{tabular}
   
    \label{tab:1}  
\end{center}
\end{table}

According to Table \ref{tab:1} we can know that PCG has low complexity under MIMO-AFDM system, then analyzing the complexity of the individual algorithms and the performance of PCG will be presented in the next section.
\section{SIMULATION RESULTS}
In this paper, we focus on the application of our proposed efficient algorithm in MIMO-AFDM systems, comparing the complexity and communication performance with BER.

\begin{figure}[t]
\centering
\includegraphics[width=0.55\textwidth]{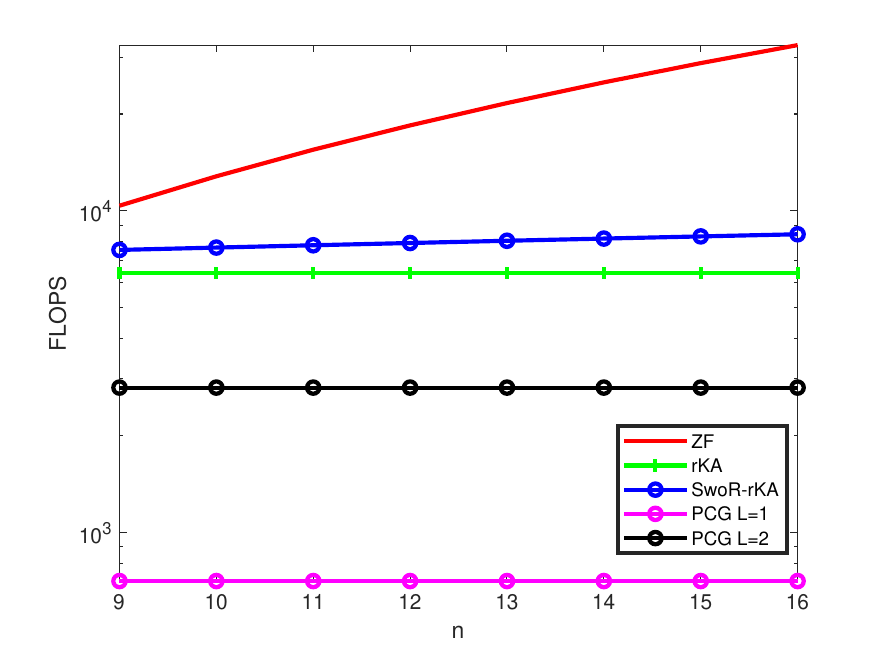}
\caption{ MIMO-AFDM of Scenario.}
\label{fig:3}
\end{figure}

In the first example, shown in \figureref{fig:3}, we can see that our proposed PCG algorithm has lower complexity and similar performance compared to ZF\cite{b12}. Consider a BS configured with $N_t = 64$ transmitting antennas broadcasting data to $K = 9-16$ users and each user configured with $N_k = 1 $ receiving antenna. 

From \figureref{fig:3}, we can see that the algorithm complexity of rKA, SworR-rKA and PCG does not change much with the increase of users, indicating that the rKA, SworR-rKA and PCG algorithm complexity is only related to the transmitting antenna array and the number of algorithm iterations, which will be more advantageous to use in scenarios with changing users. Also PCG is more advantageous in terms of complexity overhead under the condition of sparsifying the channel $\textbf H$. We can know that rKA and SworR-rKA reduce the precoding complexity overhead by algorithmic means, while we propose PCG to reduce the complexity overhead based on the AFDM channel characteristics.

\begin{figure}[t]
\centering
\includegraphics[width=0.55\textwidth]{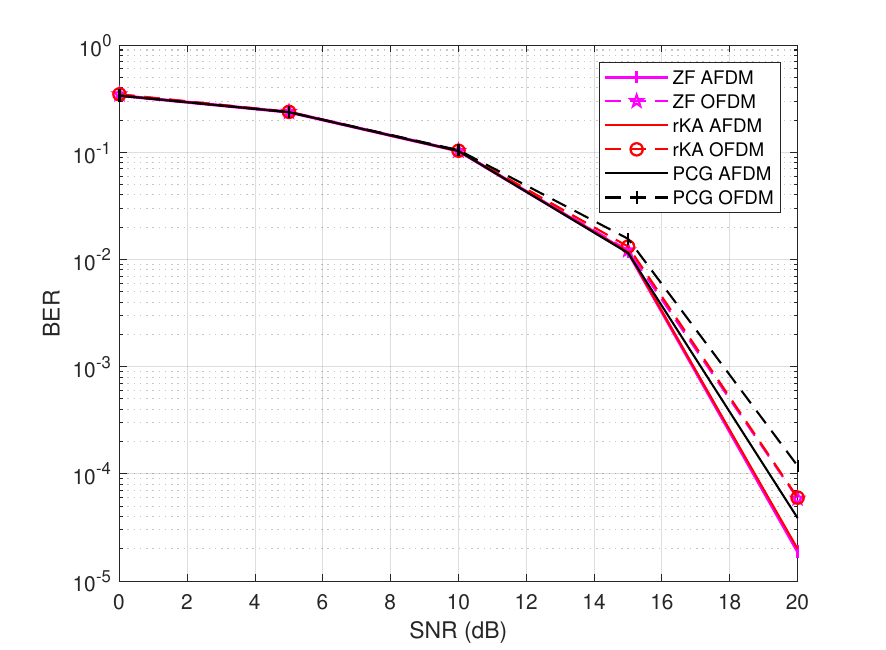}
\caption{ BER versus SNR (dB) for spatially precoded AFDM compared with OFDM$(N_t = 64,K = 16,N_k = 1)$.}
\label{fig:4}
\end{figure}

In the second example, shown in \figureref{fig:4}, we evaluate the BER of the proposed algorithm. Consider a BS configured with $N_t = 64$ transmitting antennas broadcasting data to $K = 16$ users  and each user configured with $N_k = 1 $ receiving antenna and $v_i $ =100 in \eqref{eq:4} for mobility scenario. Compared to ZF \cite{b12}, our proposed PCG in AFDM has a lower BER than OFDM. The PCG can achieve similar performance in both AFDM and OFDM compared to ZF.

Simulation results show that PCG has lower complexity and similar performance compared to the ZF of \cite{b12}. Also compared to \cite{b9}, \cite{b10}, and \cite{b11}, AFDM can show better performance compared to OFDM in mobility scenarios. Meanwhile, the BER based on PCG-AFDM  in the simulation is lower, and the performance is optimized by about 1 dB.

\section{CONCLUSION}
In this paper, we focus on precoding design under the MIMO-AFDM system model. Due to the challenges of DFO in mobile scenarios, we employ AFDM as a solution to mitigate this problem. To address both the MUI and computational complexity challenges in MIMO-AFDM systems, we propose a two-stage approach. First, we leverage the inherent channel sparsity characteristics of AFDM systems to perform element-wise selection in the channel matrix using eSNR criteria, eliminating components with negligible impact on BER. Subsequently, we apply the PCG algorithm to the sparsified system, yielding a precoding scheme with significantly reduced computational overhead.


\vspace{12pt}
\color{red}

\end{document}